\documentclass[aps,prl,preprint,groupedaddress,showpacs,superscriptaddress]{revtex4-1}

\usepackage{graphicx}
\usepackage{float,amsmath,verbatim,amssymb}
\usepackage{epsfig}

\begin{document}


\title{DC magnetic field generation in unmagnetized shear flows}

\author{T. Grismayer }
\email{thomas.grismayer@ist.utl.pt}
\author{E. P. Alves\footnote{T. Grismayer and E. P. Alves contributed equally to this work}}
\affiliation{GoLP/Instituto de Plasmas e Fus\~ao Nuclear - Laborat\'orio Associado, Instituto Superior T\'ecnico, Lisbon, Portugal}
\author{R. A. Fonseca}
\affiliation{GoLP/Instituto de Plasmas e Fus\~ao Nuclear - Laborat\'orio Associado, Instituto Superior T\'ecnico, Lisbon, Portugal}
\affiliation{DCTI/ISCTE Instituto Universit\'{a}rio de Lisboa, 1649-026 Lisboa, Portugal}
\author{L. O. Silva}
\email{luis.silva@ist.utl.pt}
\affiliation{GoLP/Instituto de Plasmas e Fus\~ao Nuclear - Laborat\'orio Associado, Instituto Superior T\'ecnico, Lisbon, Portugal}

\date{\today}

\begin{abstract}
The generation of DC magnetic fields in unmagnetized plasmas with velocity shear is predicted for non relativistic and relativistic scenarios either due to thermal effects or due to the onset of the Kelvin-Helmholtz instability (KHI). A kinetic model describes the growth and the saturation of the DC field. The predictions of the theory are confirmed by multidimensional particle-in-cell simulations, demonstrating the formation of long lived magnetic fields ($t \sim 100s~\omega_{pi}^{-1}$)  along the full longitudinal extent of the shear layer, with transverse width on the electron length scale ($\sqrt{\gamma_0}c/\omega_{pe}$), reaching magnitudes $eB_{\mathrm{DC}}/m_ec\omega_{pe}\sim \beta_0\sqrt{\gamma_0}$.
\end{abstract}



\maketitle

It is now recognized that kinetic plasma instabilities, operating on the electron time scale in unmagnetized plasmas, such as the Weibel instability \cite{weibel59}, can be of importance to explain the generation of magnetic fields relevant for shock formation in astrophysical and laboratory conditions and for particle acceleration and radiation emission in astrophysical scenarios \cite{medvedev99,gruzinovwaxman99}. Ab initio particle-in-cell simulations (PIC) have demonstrated the generation of subequipartition magnetic fields, i.e., the ratio between the energy density of the field and the kinetic energy density of the flow are close to $10^{-3} - 10^{-2}$ \cite{silva03,spitkovsky08,frederiksen04}. Previous studies have not considered the role of velocity shears in the dynamics of unmagnetized plasmas, which is known to lead to the onset of the collisionless Kelvin-Helmholtz instability (KHI). This scenario is of relevance from several perspectives: it has been proposed that the magnetic dynamo can operate over the seed magnetic field due to the collisionless KHI \cite{gruzinov08,macfadyen09}, it has been suggested that the KHI could play an important part in secondary magnetic island generation in magnetic reconnection sites \cite{Fermo12} and the KHI remains a well known benchmark of magnetohydrodynamic simulations \cite{mignone09,beckwith11}. However the dynamics on the electron time-scale of the KHI is unexplored, and recent experiments have now addressed the collisional KHI \cite{harding09,kuramitsu12,Hurricane12} and in the near future, with the advent of more powerful lasers, experiments will be able to probe the collisionless KHI, in scenarios relevant for astrophysics. 
Motivated by recent particle-in-cell simulations that have demonstrated the onset of large scale DC magnetic fields in unmagnetized plasmas with velocity shear \cite{alves12}, not captured by two-fluid theory \cite{dangelo65,gruzinov08,alves12}, we develop a theoretical model to predict the onset, growth and saturation of this magnetic field, as well as its spatial and long time behavior features. 
We compare the theoretical predictions with one (1D), two (2D) and three (3D) dimensional PIC simulations. 

We recall the 2D theoretical model of the unmagnetized KHI \cite{gruzinov08,alves12}, which is based on the relativistic fluid formalism of plasmas coupled with Maxwell's equations. Without loss of generality, we focus on symmetrically shearing flows (with velocities $\pm v_0\vec{e_y}$ along the y direction and with equal densities $n_0$) with a tangential discontinuity in the x direction. The protons are considered free-streaming whereas the electron fluid quantities and fields are linearly perturbed, $u=\bar{u}~e^{-k_\perp |x|}e^{i(k_\parallel y-\omega t)}$. The unstable modes are stationary ($\mathrm{Re}(\omega)=0$) surface waves and obey the following dispersion relation:
\begin{equation}
\frac{\Gamma}{\omega_{pe}} = 
	\left[
		\frac{1}{2\gamma_0^3} \left(
			\sqrt{1+8 \frac{k_\parallel^2v_0^2\gamma_0^3}{\omega_{pe}^2}}-1-2\frac{k_\parallel^2v_0^2\gamma_0^3}{\omega_{pe}^2} 
		\right) 
	\right]^{1/2},
\end{equation}
where $\Gamma$ is the growth rate ($\mathrm{Im}(\omega)$) of the mode with wave number $k_\parallel$, $\omega_{pe}=\sqrt{4\pi n_0e^2/m_e}$ is the plasma frequency, $k_\perp^2= k_\parallel^2+\omega_{pe}^2/(c^2\gamma_0^3)-\omega^2/c^2$ and $\gamma_0=1/\sqrt{1-\beta_0^2}$ is the Lorentz factor of the shearing flows with $\beta_0=v_0/c$.  After a few e-folding times the system is dominated by the fastest growing mode with $\Gamma_\mathrm{max}=\sqrt{1/8}\gamma_0^{-3/2}~\omega_{pe}$ and $k_{\parallel \mathrm{max}}=\sqrt{3/8}\gamma_0^{-3/2}~\omega_{pe}/v_0$. 

To ascertain these results and to fully explore the KHI, particle-in-cell (PIC) simulations were performed using OSIRIS \citep{fonseca02,fonseca08}. We simulate shearing slabs of cold ($v_{0} \gg v_{th}$, where $v_{th}$ is the thermal velocity) unmagnetized electron-proton plasmas with a reduced mass ratio $m_p/m_e=100$ ($m_e$ and $m_p$ are respectively the electron and the proton mass), and evolve it up to $\omega_{pi}t =100$ ($\omega_{pi}=\sqrt{m_e/m_p}~\omega_{pe}$ is the proton plasma frequency). First we present the simulation results of a shear flow with $v_0=0.2c$. The shear flow initial condition is set by a velocity field with $+v_0\vec{e_y}$ in the middle-half of the simulation box, and a symmetric velocity field with $-v_0\vec{e_y}$ in the upper and lower quarters of the box. Initially, the system is charge and current neutral. Periodic boundary conditions are imposed in every direction. The simulation box dimensions are $500 \times 200  ~ ( c/\omega_{pe} )^2$, resolved with $20$ cells per electron skin depth ($c/\omega_{pe}$), and a number of 36 particles per cell per species are used. Space and time are respectively normalized to $c/\omega_{pe}$ and $1/\omega_{pe}$.

\begin{figure}[t!]
\begin{center}
\includegraphics[width=1\columnwidth]{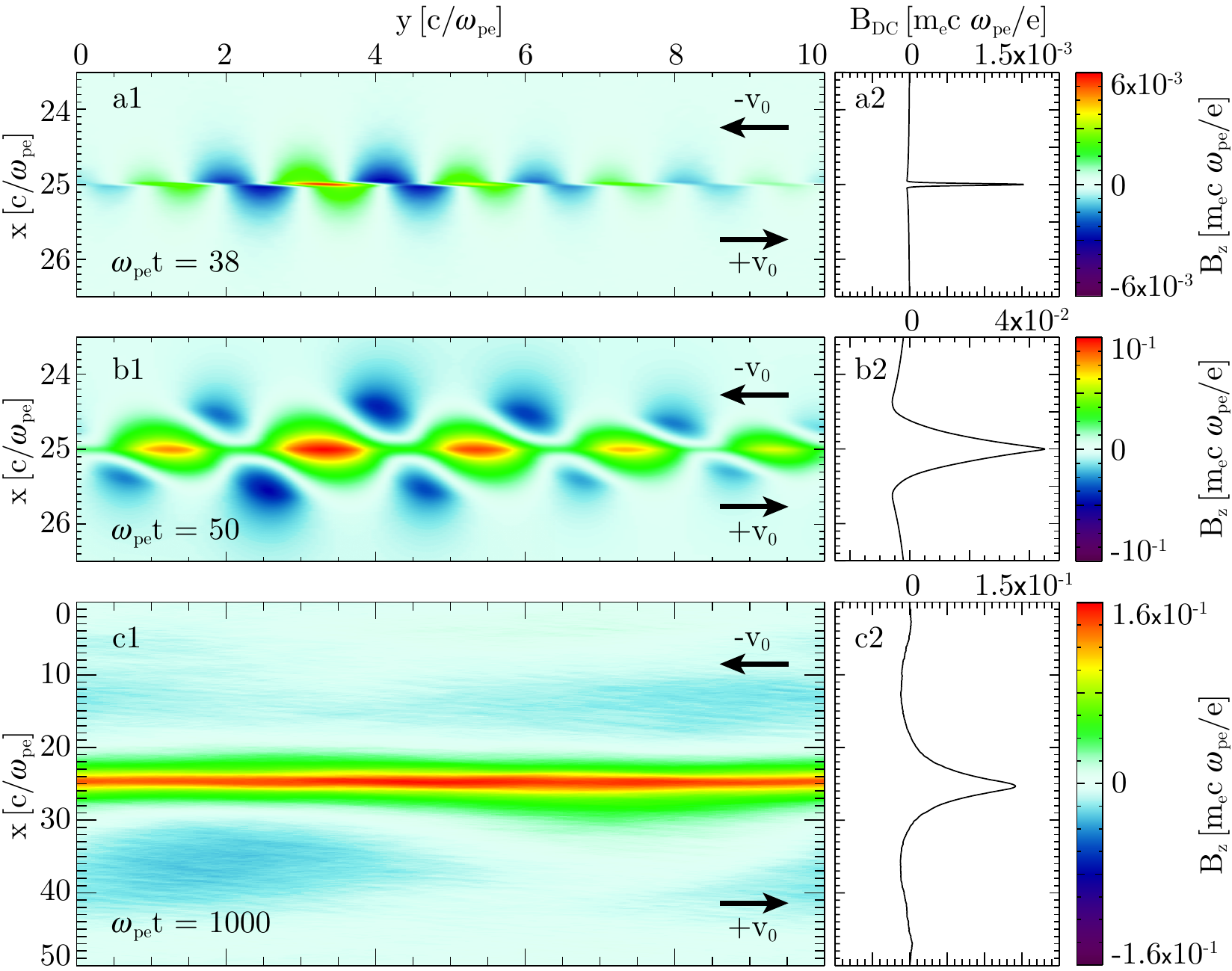}
\caption{\label{fig:KHI2d} $B_z$ component of the magnetic field structure generated by the cold KHI for $v_0=0.2c$ during (a) the linear regime, (b) near saturation and (c) at $t=1000~\omega_{pe}^{-1}=100~\omega_{pi}^{-1}$. The insets on the right hand side represent the longitudinal average of the magnetic field, revealing the DC component.}
\end{center}
\end{figure}

The growth and the wavenumber of the most unstable mode are confirmed by 2D simulations (Fig. \ref{fig:KHI2d} (a1)). Interestingly, the growth of a DC ($k_\parallel=0$) magnetic field mode is also observed (insets (a2) and (b2) of Fig. \ref{fig:KHI2d}), which is not predicted by the linear fluid theory ($\Gamma(k_\parallel=0)=0$) nor has it been previously identified in MHD simulations \cite{macfadyen09,Frank96,ChenHasegawa74,ZhuKivelson88} and only recently in kinetic simulations \cite{alves12}. The growth of the DC magnetic field mode results from a current imbalance due to the mixing between the electron flows across the shear surface, while the proton flows remain almost unperturbed due to their inertia. 
The mixing arises due to the deformation of the electron interface between the two flows, which in the linearized fluid calculations is not accounted for and, in zeroth order, remains fixed. Alternatively, we find that the physics describing the formation of a DC mode can be modeled in a 1D reduced theory where an initial temperature drives the mixing effect.
In order to understand the formation of the DC field, we first consider the one-dimensional case. Initially all the fields are zero and we assume a plasma with a tangential shear flow in the $x$ direction and an initial temperature such that $v_{th}\ll v_0$. It is clear that the thermal expansion of the electrons across the shear surface (ions are assumed to be cold and free streaming) leads to an imbalance of the current neutrality around the shear surface, forming a DC magnetic field in z direction. The initial corresponding electron distribution function reads $f(x,v_x,v_y,v_z,t=0)=f_0(v_x,v_y-v_0~\mathrm{sgn}(x),v_z)$. Due to the dimensionality of the problem, it is clear that $E_z, B_x, B_y$ remain zero. The reduced set of equations is therefore Maxwell's equations coupled with the Vlasov equation $\partial_t F + v_x\partial_x F-e/m_e(\vec{E}+(\vec{v}/c)\times\vec{B_z}).\partial_{\vec{v}} F=0$ where $F(x,v_x,v_y,t)=\int dv_z f(x,v_x,v_y,v_z,t)$. The formal solution of the Vlasov equation is $F(x,v_x,v_y,t)=F_0(x_0,v_{x0},v_{y0})$, where $x_0$, $v_{x0}$ and $v_{y0}$ denote the position and velocities of an electron at $t=0$ and $F_0=\int dv_{z0}f_0$ \cite{ONeil,Morales72}. At early times, if we assume that the induced fields are sufficiently small that we can neglect their effect on the change of momentum of the electrons, the distribution function can be determined along the free streaming orbits of the electrons \cite{Grismayer11}. For the sake of simplicity, we separate the initial electronic distribution in two parts, $F_0=F_0^{-}(x_0<0)+F_0^{+}(x_0>0)$. The electron currents read $J_{e,y}^{\pm}\simeq-e\int dv_{y}v_{y}\int dv_{x} F_0^{\pm}(x-v_xt,v_x,v_y\mp v_0)$. 
For a Maxwellian distribution function, $f_M(v)=e^{-v^2/2v_{th}^2}/\sqrt{2\pi} v_{th}$, we have $F_0^{\pm}(x_0,v_{x0},v_{y0}\mp v_0)=n_0f_M(v_{x0})f_M(v_{y0}\mp v_0)$ and we obtain $J_{e,y}^{\pm}\simeq \mp en_0v_0  \int_{\mp x/t}^{\infty} dv_{x} f_M(v_x) \simeq \mp ev_0n_0~\mathrm{erfc}\left(\frac{\mp x}{\sqrt2v_{thx}t}\right)$. The total current is given by adding the unperturbed proton currents $J_{p,y}=en_0v_0\mathrm{sng}(x)$, from which the magnetic field can be integrated by neglecting the displacement current in Amp\`ere's Law:
\begin{eqnarray}
B_\mathrm{DC}&\simeq& e4\pi n_0\beta_0\sqrt{2}v_{thx}t\left[\frac{e^{-\xi^2}}{\sqrt{\pi}}-\xi~\mathrm{erfc}(\xi)\right],
\label{bdcwarm}
\end{eqnarray}
where $\xi=|x|/\sqrt{2}v_{thx}t$. We thus verify that the DC magnetic field driven by the electron thermal expansion grows linearly with time. Its typical width is on the order of $\sqrt{2}v_{thx}t$ and its peak intensity $B_\mathrm{DC}(x=0)=4\sqrt{2\pi}en_0\beta_0v_{thx}t$. This derivation is valid as long as the orbits of the electrons do not diverge much from the free streaming orbits, i.e., as long as the electric and magnetic fields that develop self-consistently do not affect the free motion of the particles. However, the electrons will eventually feel the induced magnetic field which tends to push more electrons across the shear via the $\vec{v}_0 \times \vec{B}_\mathrm{DC}$ force. Consequently, the rate at which electrons cross the shear increases, which, in turn, enhances the growth rate of the magnetic field.

\begin{figure}[t]
\begin{center}
\includegraphics[width=1.0\columnwidth]{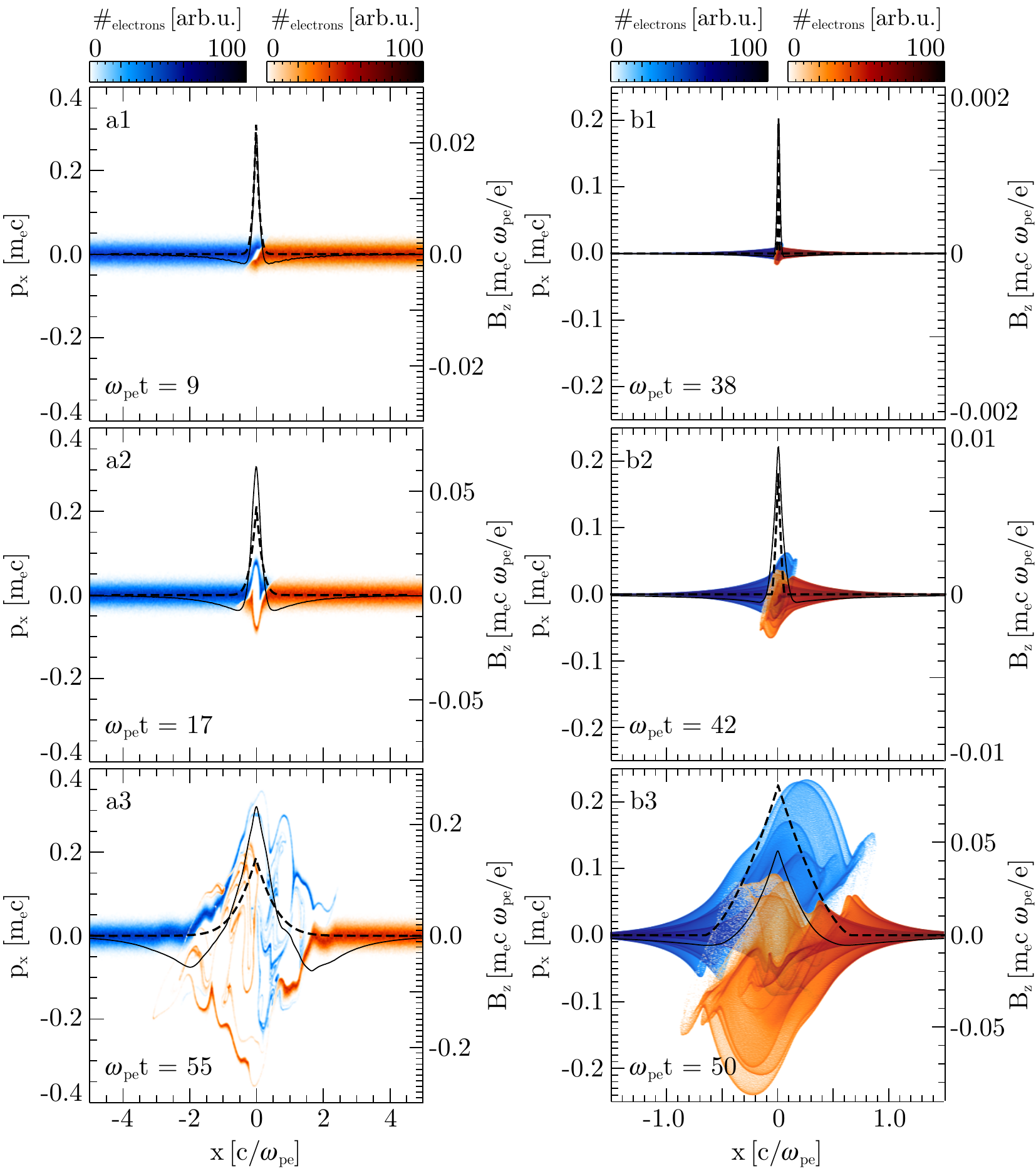} 
\caption{\label{fig:p1x1}
Evolution of the electron phasespace. Left: 1D warm shear flow with $v_0=0.2~c$ and $v_\mathrm{th}=0.016$. Right: 2D cold shear flow with $v_0=0.2c$. The blue (red) color represents the electrons with a negative (positive) drift velocity $v_0$. The self-consistent DC magnetic field is represented by the solid curve, whereas the dashed curve represents the magnetic field given by the theoretical model.}
\end{center}
\end{figure}

\begin{figure}[t]
\includegraphics[width=0.8\columnwidth]{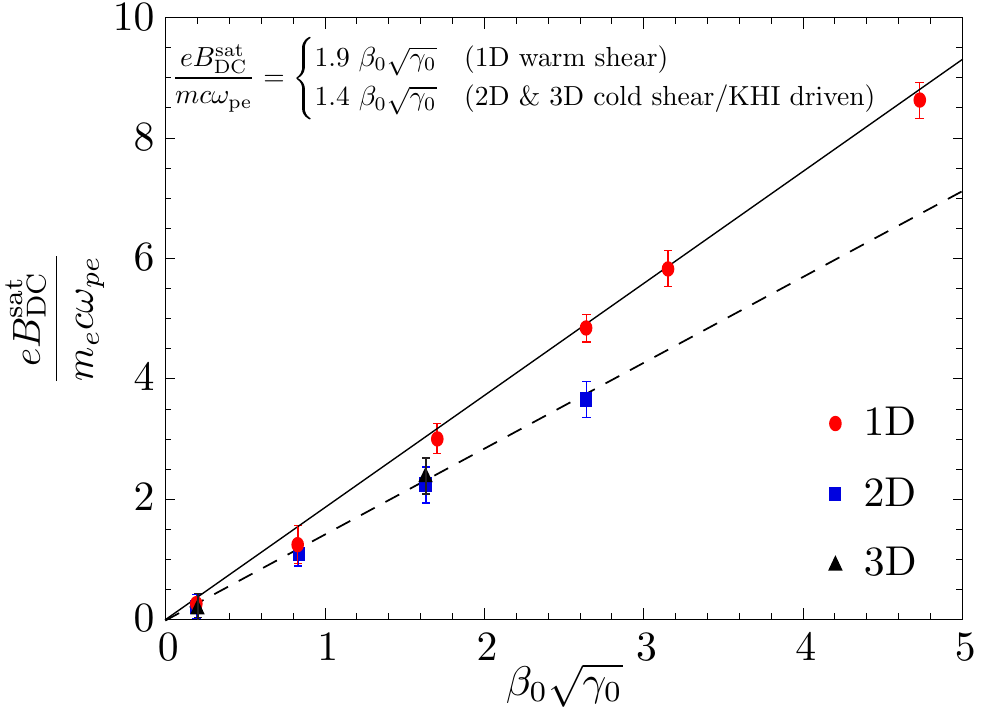}
\caption{\label{Bdcscaling} Magnitude of the DC magnetic field peak at saturation as a function of $\beta_0\sqrt{\gamma_0}$. The red, blue and black markers represent the results of 1D, 2D and 3D PIC simulations, respectively. The error bars are associated to the fluctuations of the peak value in the saturation stage. The lines represent best fit curves to the simulation results, demonstrating the good agreement with Eq.~\ref{eq:bsat}.}
\end{figure}

In order to verify our analytical calculations and to further investigate the phase where electrons deviate from their free streaming orbits, we carried out 1D simulations of the $x$ direction of the 2D simulations. The Debye length is resolved in the 1D simulations ($\Delta x = \lambda_D$) and we used 1000 particles per cell. Fig. \ref{fig:p1x1} (a1-3) shows the time evolution of the $x p_x$ phase space and the magnetic field for $v_{th}=0.016 c$ and $v_0=0.2c$. At earlier times $\omega_{pe}t=9$ (Fig. \ref{fig:p1x1} (a1)) an excellent agreement between the model and the simulation is observed.
The model breaks down approximately when an electron initially with $v_x=0$ (around the shear) acquires a velocity change on the order of $v_{thx}$, which corresponds to a strong distortion of the Maxwellian distribution around the shear. 
Fig. \ref{fig:p1x1} (a2) shows this effect at $\omega_{pe} t=17$. The model underestimates the magnitude of the magnetic field and one can clearly observe the distortion of the distribution function in the field region. As the magnetic field grows, the Larmor radius ($r_L$) of the electrons crossing the shear interface decreases. When the minimum $r_\mathrm{L,min}$ (associated to the peak of $B_\mathrm{DC}$) becomes smaller than the characteristic width of the magnetic field $l_\mathrm{DC}$, the bulk of the electrons becomes trapped by the magnetic field structure. This is illustrated in Fig. \ref{fig:p1x1} (a3) at $\omega_{pe} t= 55$. The magnetic trapping prevents the electron bulk expansion across the shear (that drives the growth of the magnetic field), saturating the magnetic field. An estimate of the saturation can be obtained by equating $r_\mathrm{L,min} \sim l_\mathrm{DC}$. From Eq.~(\ref{bdcwarm}), it is possible to write the magnetic field as $B_\mathrm{DC}(x,t)=4\pi en_0\beta_0w(x,t)$, where $w(0,t)$ should be interpreted as the characteristic width of the field. With $l_\mathrm{DC}\sim w(0,t)$, $r_\mathrm{L,min}=mv_0\gamma_0/em_eB_\mathrm{DC}(0,t)$, we find that $l_\mathrm{DC}\sim c\sqrt{\gamma_0}/\omega_{pe}$ giving the saturation level of the magnetic field as $eB_\mathrm{DC}^\mathrm{sat}/m_ec\omega_{pe}\sim \beta_0\sqrt{\gamma_0}$.
This scaling has been verified for 1D simulations (Fig. \ref{Bdcscaling}).


In the absence of an initial temperature, an alternative mechanism is needed to drive the electron mixing across the shear surface that in turn generates the DC field. This mechanism is the cold fluid KHI that operates in 2D and 3D geometries. In fact, in the warm shear flow scenario, both the cold fluid KHI and the electron thermal expansion can contribute to the generation of the DC field. This happens when the typical length of the DC field due to the thermal expansion ($l_\mathrm{DC}$) after a few e-foldings of the cold fluid KHI ($T_\mathrm{KHI-growth} = n_\mathrm{e-foldings}/\Gamma_\mathrm{max}$, where $n_\mathrm{e-foldings}$ is on the order of 10) is on the order of the relativistic electron skin depth, i.e., $v_{th} T_\mathrm{KHI-growth} \sim \sqrt{\gamma_0}c/\omega_{pe}$. Therefore, the cold fluid KHI dominates the electron mixing in the limit  $v_{th} T_\mathrm{KHI-growth} \ll \sqrt{\gamma_0}c/\omega_{pe}$.

For a two dimensional cold plasma undergoing the KHI, the electron distribution function can be written as $f(x,y,v_x,v_y,v_z,t)=n_0\delta(v_x-v_{xfl}(x,y,t))\delta(v_y-v_{yfl}(x,y,t))\delta(v_z)$ where $v_{xfl},v_{yfl}$ correspond to the velocity field solutions of the fluid theory. 
In this case, the self-generated KHI fields play the role of an effective temperature that transports the electrons across the shear surface, while the protons remain unperturbed, inducing a DC component in the current density, and hence in the fields. We then have to solve the evolution of the distribution function and show that the current density $J_y$, averaged over a wavelength $\lambda=2\pi/k_\parallel$, has a non zero DC part. We follow the same approach as before and calculate the average distribution function defined as:
\begin{equation}
\label{eq:fave}
F(x,v_x,t)=\frac{1}{\lambda}\int dv_y\int dv_z \int_{\lambda} dy f(x,y,v_x,v_y,v_z,t).
\end{equation}
To obtain analytical results we will assume that the linearly perturbed fluid quantities are purely monochromatic, which is equivalent to assume that after a few e-foldings, the mode corresponding to $k_\parallel=k_{\parallel \mathrm{max}}$ dominates with a growth rate of $\Gamma = \Gamma_{max}$. We then write $v_{yfl}\simeq v_0(x)$ and $v_{xfl}=\bar{v}_{xfl}\sin(k_\parallel y)e^{-k_\perp|x|+\Gamma t}$, where $\bar{v}_{xfl}$, the amplitude of the velocity perturbations at $t=0$, is associated to the small thermal fluctuations (small enough to ensure that the thermal expansion is negligible over $T_\mathrm{KHI-growth}$). Inserting $v_{xfl},v_{yfl}$ into Eq.~(\ref{eq:fave}), we obtain
\begin{eqnarray}
F(x,v_x,t)=\frac{n_0}{\pi v_{max}\sqrt{1-\xi^2}},
\end{eqnarray}
where $\xi(x,v_x,t)=v_x/v_{max}(x,t)$ with $v_{max}(x,t)=\bar{v}_{xfl}e^{-k_{\perp}|x|+\Gamma t}$.
We observe that the development of 2D cold KHI reveals close similarities with the 1D hot model previously described. In the 2D KHI, averaging the distribution in the direction of the flow shows that the perturbation gives rise to a spread in $v_x$ that may be interpreted as an effective temperature. The spread in $v_x$ decays exponentially away from the shear and grows exponentially with time. The mean velocity is zero and the effective temperature associated to this distribution function is defined as $V_{eff}^2(x,t)=(1/n_0)\int dv_x v_x^2 F(x,v_x,t)=v_{max}^2/2$. One can then expect a similar physical picture as in the hot shear scenario and, as a result, the emergence of DC components in the fields which are induced by the development of the unstable KH perturbations. The evolution of the phase space in Fig. \ref{fig:p1x1} illustrates the similarity between the warm 1D (insets a1-3) and cold 2D (insets b1-3) scenarios. 

The challenge in this scenario is to determine how such a distribution function expands across the shear surface due to the complexity of the orbits in the fields structure (multidimensional fields with discontinuities at $x=0$). 
In the region where the electron mixing occurs, we assume electron orbits given by $x\sim x_0+(v_{x0}/\Gamma) e^{\Gamma t}$ and $v_x\sim v_{x0}e^{\Gamma t}$ where $x_0$ and $v_{x0}$ are the position and velocity of a particle at the time $t_0$ when the instability begins. 
\begin{eqnarray}
\label{eq:jdc_int}
J_{e,y}^{\pm}(x,t)&\simeq&\mp ev_0\int_{\mp x\Gamma}^{v_{max}^0} dv_x F(x,v_x,t) \\
&\simeq&ev_0n_0\left[\frac{1}{2}\pm\frac{1}{\pi}\arcsin\left(\frac{x\Gamma}{v_\mathrm{max}^0}\right)\right]
\end{eqnarray} 
where $x\Gamma \in [-v_\mathrm{max}^0,v_\mathrm{max}^0]$ and $v_\mathrm{max}^0(t)=v_\mathrm{max}(x=0,t)$ that represents the maximum velocity of a particle that was originally in the vicinity of the shear. The limits of the integral (\ref{eq:jdc_int}) represent the deformation of the boundary between the two flows on a characteristic distance of $v_\mathrm{max}^0/\Gamma$ as the instability develops. In the fluid theory, the boundary remains fixed, precluding the development of the DC mode. We then find the total current density by summing the proton contribution and integrate to obtain the induced DC magnetic field:
\begin{eqnarray}
\label{bdc_cold_expr}
B_\mathrm{DC}(x\gtrless 0,t)=\mp 4\pi en_0\beta_0\left[\mp x\left(1\mp\frac{2}{\pi}\arcsin\left(\frac{\Gamma x}{v_\mathrm{max}^0(t)}\right)\right) +\frac{2}{\pi}\sqrt{\left(\frac{v_\mathrm{max}^0(t)}{\Gamma}\right)^2-x^2}\right]
\end{eqnarray}
The peak of the DC magnetic field is located at $x=0$ where the expression above reduces to $B_\mathrm{DC}(0,t)=8e\beta_0n_0v_\mathrm{max}^0(t)/(\pi\Gamma)$ and thus grows at the same rate as the KHI fields. One can verify in Fig. \ref{fig:p1x1} (b1-b2) that Eq.~(\ref{bdc_cold_expr}) shows reasonable agreement with the 2D simulations. This derivation neglects the DC Lorentz force on the electron trajectories, which makes this model valid as long as induced DC field remain small compared to the fluid fields associated to the mode $k_{\parallel \mathrm{max}}$. The peak of $B_\mathrm{DC}$ field is proportional to $v_\mathrm{max}^0(t)$.
We therefore conclude that the induced DC magnetic field is always on the same order of the fluid fields (Fig. \ref{fig:KHI2d} (a) and (b)) and thus its consequences to KHI development cannot be neglected. As the DC field evolves, electrons start to get trapped and we expect a level of saturation similar to the one obtained in the 1D model which is verified by the simulations. The comparisons between the saturation level of the 1D, 2D and 3D simulations are shown in Fig. \ref{Bdcscaling}, verifying the $\beta_0\sqrt{\gamma_0}$ scaling. 
Interestingly, the DC magnetic field remains stable beyond the electron time scale and persists up to 100s $\omega_{pi}^{-1}$ as shown in Fig. \ref{fig:KHI2d} (c). Eventually the protons will drift away from the shear surface due to the magnetic pressure, broadening the DC magnetic field structure and lowering its magnitude. 

In conclusion, we have presented an analytical description of the formation of a DC magnetic field in a shear flow scenario, which is in good agreement with 1D, 2D and 3D PIC simulations. We have shown that the DC magnetic field can arise due to an electron thermal expansion across the shear, in a warm shear scenario; we also extended this picture to the cold shear scenario 
where the development of the cold fluid KHI perturbations induce an effective temperature that drives the electron mixing across the shear. The DC magnetic field saturates on the electron time scale and persists up to proton time-scales, reaching maximum magnitudes of $eB_\mathrm{DC}/mc\omega_{pe}\simeq 1.4\beta_0\sqrt{\gamma_0}$ with thicknesses of a few $\sqrt{\gamma_0}c/\omega_{pe}$, and thus is dynamically relevant for the evolution of the KHI on ion time scales.

This work was partially supported by the European Research Council ($\mathrm{ERC-2010-AdG}$ Grant 267841) and FCT (Portugal) grants SFRH/BD/75558/2010, SFRH/BPD/75462/2010, and PTDC/FIS/111720/2009. We would like to acknowledge the assistance of high performance computing resources (Tier-0) provided by PRACE on Jugene based in Germany. Simulations were performed at the IST cluster (Lisbon, Portugal), and the Jugene supercomputer (Germany).


\newpage

\end{document}